\documentclass[hyper,letterpaper,notoc]{JHEP3}

\usepackage[dvips]{graphicx}
\usepackage{epsfig}
\usepackage{amsfonts,amssymb}
\usepackage[dvips]{graphicx}

\title{One-Loop Corrections to the $\rho$ Parameter in Higgsless Models}

\author{Baradhwaj Coleppa and Stefano Di Chiara\\
Department of Physics and Astronomy, Michigan State University\\
East Lansing, MI 48824, USA\\
E-mail: baradhwa@msu.edu, dichiara@msu.edu}

\author{Roshan Foadi\\
Niels Bohr Institute, Copenhagen, DK-2100, \\ 
and Southern Denmark University, Odense, DK-5000 \\
Denmark \\
E-mail: roshan@fysik.sdu.dk}

\abstract{A large class of deconstructed Higgsless model is known to satisfy the tree-level experimental bounds on the electroweak precision parameters. In particular, an approximate custodial symmetry insures that the tree-level $\rho$ parameter is exactly one, for arbitrary values of the model parameters, and regardless of fermion delocalization. In this note we expand on previous work by considering the fermionic one-loop contributions to $\rho$, which are essentially due to loops with top and bottom modes. We analyze the dependence on the number $N$ of internal SU(2) sites in models with a ``flat background''. We find that the new-physics contribution rapidly increases with $N$, to quickly stabilize for large values of $N$. Experimental upper bounds on $\rho$ translate into lower bounds on the mass of the heavy fermions. These, however, are weakly correlated with $N$, and the three-site model ($N=1$) turns out to be already an excellent approximation for the continuum model ($N\rightarrow\infty$).}


\preprint{}

\begin{document}

\section{Introduction}
Higgsless models~\cite{Csaki:2003dt} are effective field theories which break the electroweak symmetry without producing a scalar Higgs boson. The most popular among these models are based on an SU(2)$\times$SU(2)$\times$U(1) gauge theory on a slice of AdS$_5$ space~\cite{Agashe:2003zs}~\cite{Csaki:2003zu}, where the electroweak symmetry is broken by an appropriate choice of boundary conditions. The five-dimensional gauge fields can be expanded in Kaluza-Klein (KK) towers of four-dimensional charged and neutral vector bosons, where the lightest modes correspond to the ordinary electroweak gauge bosons (including the massless photon). Longitudinal $W$ and $Z$ boson scattering amplitudes are then unitarized through exchanges of the massive modes~\cite{SekharChivukula:2001hz}~\cite{Chivukula:2002ej}~\cite{Chivukula:2003kq}~\cite{He:2004zr}, where a non-zero background warping factor insures that the mass of the lightest of these heavy modes is pushed above the current lower bounds imposed by direct searches~\cite{Csaki:2003zu}.  Other extra-dimensional models employ flat backgrounds, with brane kinetic terms simulating the effects of warping~\cite{Barbieri:2003pr}~\cite{Foadi:2003xa}. Higgsless models can also be studied in a four-dimensional context by using the technique of deconstruction~\cite{Arkani-Hamed:2001ca}~\cite{Cheng:2001vd}, or without any referral to extra-dimensions, by constructing the most general chain of non-linear sigma models with arbitrary gauge couplings and $f$-constants~\cite{Chivukula:2004pk}-\cite{Casalbuoni:2005rC}.

Most of the recent efforts on Higgsless physics have been focused on the tension between unitarity, which demands the new vector bosons to be relatively light, and electroweak precision data, which instead favor heavy vector bosons. Clearly the corrections to electroweak observables depend crucially on the way matter fields are coupled to the gauge sector of the model. The simplest choice is to have fermions strictly localized at the ends of the extra-dimensional interval. With this choice no extra fermions are introduced into the model, but it turns out to be impossible to simultaneously satisfy the experimental constraints on the Peskin-Takeuchi $S$ and $T$ parameters~\cite{Foadi:2003xa}~\cite{Chivukula:2004pk} and unitarity.  It is therefore necessary to delocalize fermions, or, in other words, to have five-dimensional matter fields propagating into the bulk of the extra dimension. As for the gauge fields, this introduces towers of four-dimensional fermions, with the lowest mode of each tower corresponding to a standard model fermion. The latter couples not only with the gauge fields at the interval ends, but also with the bulk gauge fields. However, if the profile of the left-handed light fermions is adjusted to mimic the profile of the standard model W boson, then the former will be ``orthogonal'' to the heavy charged vector bosons, and the corresponding couplings may vanish, decoupling the light fermions from the new physics. This has been proved to be possible in a large class of four-dimensional Higgsless models, consisting of an SU(2)$^{N+1}\times$U(1) chain of non-linear sigma models with arbitrary parameters, where three of the four leading zero-momentum electroweak parameters defined by Barbieri {\em et.al}~\cite{Barbieri:2004qk} can indeed be simultaneously adjusted to exactly vanish~\cite{SekharChivukula:2005xm}. In models from extra dimension, an exact vanishing of all the electroweak parameters is not possible, since the profile of left-handed fermions cannot be shaped arbitrarily. However the $S$ parameter can be tuned to zero, and all other parameters are naturally suppressed~\cite{Cacciapaglia:2004rb}~\cite{Foadi:2004ps}~\cite{Casalbuoni:2005rC}.

In this paper we focus on the contribution to one of the electroweak parameters, namely the $\rho$ parameter, defined as the ratio between the strengths of the isotriplet neutral current and charged current interactions at zero momentum. At tree level this computation has been done for the SU(2)$^{N+1}\times$U(1) model with arbitrary parameters, for which it has been proved in an elegant way that $\rho=1$ exactly, regardless of fermion delocalization~\cite{SekharChivukula:2005xm}. In fact this is achieved quite naturally thanks to an approximate custodial isospin symmetry, which becomes exact when hypercharge and Yukawa interactions are turned off. One-loop contributions to $\rho$ in a simple three-site model (corresponding to $N$=1) are calculated in Ref.~\cite{SekharChivukula:2006cg} for fermionic loops, and in Ref.~\cite{Matsuzaki:2006wn} for loops with gauge and Goldstone bosons. The latter give cutoff dependent contributions, reflecting the fact that Higgsless models are non- renormalizable effective theories of electroweak symmetry breaking. However the fermionic loop contributions are free of infinities, and thus phenomenologically relevant. We therefore focus on these by extending the corresponding analysis to models with an arbitrary number of sites and a ``flat background''. By this we mean index-independent parameters for the internal sites and links, and arbitrary parameters for the end sites. In these computations we  only consider loops from the third generation of KK quarks, since these are the only ones which give non-negligible contributions. We observe that the new-physics contribution to $\Delta\rho\equiv\rho-1$ rapidly increases to quickly stabilize, as the number $N$ of internal sites increases. The experimental upper bounds on $\Delta\rho$ translate into lower bounds for the mass of the heavy fermions. These, however, turn out to be very weakly correlated to $N$. The bounds from $\Delta\rho$ turn out to be stronger than the ones imposed by the top quark mass and the decay $b\rightarrow s+\gamma$, which were considered in previous works~\cite{SekharChivukula:2006cg}~\cite{Foadi:2005hz}.

This paper is organized as follows. In section~\ref{sec:general-model} we briefly review the most general SU(2)$^{N+1}\times$U(1) Higgsless model, with arbitrary gauge couplings, $f$-constants, Yukawa couplings, and Dirac mass terms. We give formal expressions for the tree-level low energy effective Lagrangians in terms of propagator matrix elements, and derive an expression for the $\rho$ parameter. In section~\ref{sec:one-loop} we include radiative corrections, and compute one-loop fermionic contributions from the top and bottom KK modes in the models with a flat background. We show analytical results for $N=1$ and $N\rightarrow\infty$ (corresponding to a continuum theory space model), and numerical results for arbitrary $N$, arguing that the infinities cancel in each case. In section~\ref{sec:experimental} we compare these results with the experimental bounds on $\Delta\rho$, which translate into lower bounds on the heavy fermion masses. Finally, in section~\ref{sec:conclusions} we offer our conclusions.

\section{The Model and Electroweak Interactions}\label{sec:general-model}

The Higgsless theories we consider in this note are SU(2)$^{N+2}$ non-linear sigma models whose SU(2)$^{N+1}\times$U(1) part is gauged. In Ref.~\cite{SekharChivukula:2005xm} this class of models was studied, at tree-level, in its most general form, with arbitrary parameters. The corresponding moose diagram is shown in Fig.~1. To leading order, the gauge sector Lagrangian is
\begin{eqnarray}
{\cal L}_{\rm gauge}=-{1\over 4}\sum_{j=0}^{N+1}{1\over g_j^{2}}W'^a_{j\mu\nu}W'^{a\mu\nu}_j
+{1\over 4}\sum_{j=1}^{N+1} f_j^2{\textrm tr}\left(D_\mu U_j\right)^\dagger D^\mu U_j\ ,
\label{eq:gauge-lagrangian}
\end{eqnarray} 
where
\begin{eqnarray}
D_\mu U_j=\partial_\mu U_j-i W'^a_{(j-1)\mu}T^a U_j+i U_j W'^a_{j\mu} T^a\ .
\label{eq:covariant}
\end{eqnarray}
Here $T^a=\sigma^a/2$, $a=1,2,3$, where $\sigma^a$ are the Pauli matrices. Since the last site is a U(1) gauge group, we have $W'^1_{(N+1)\mu}=W'^2_{(N+1)\mu}=0$, and the corresponding field-strength tensor is the usual Abelian one. As the link fields acquire their vacuum expectation value, the last term in Eq.~(\ref{eq:gauge-lagrangian}) becomes a mass term for the gauge bosons. The spectrum consists of $N+1$ massive charged bosons, $N+1$ massive neutral bosons, and a massless photon. Diagonalizing the mass matrices for the charged sector and the neutral sector gives expressions for the gauge eigenstates in terms of the mass eigenstates,
\begin{eqnarray}
W'^\pm_j &=& \sum_{n=0}^N a_{jn} W^\pm_n \nonumber \\
W'^3_j &=& e A + \sum_{n=0}^N b_{jn} Z_n \ , 
\label{eq:gauge-expansion}
\end{eqnarray}
where $A$, $Z_0$, $W^\pm_0$ correspond to the electroweak gauge bosons. Notice that the coefficient of the photon field is necessarily $e$, because the latter is the gauge boson of an unbroken symmetry, and must couple to any field with its coupling strength.

\EPSFIGURE[!t]{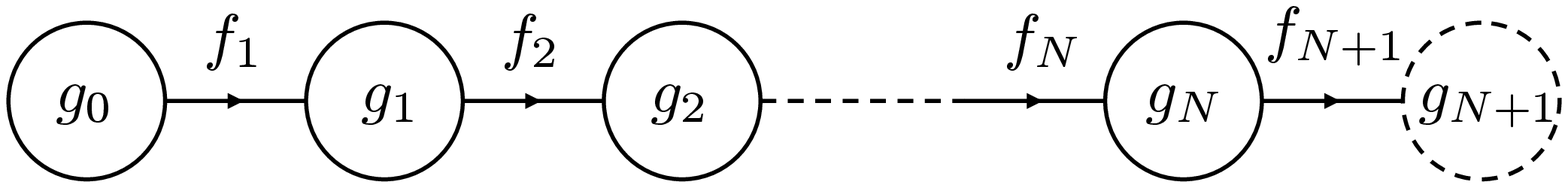,width=14cm,height=2cm}{Moose diagram for the model analyzed in Ref.~\cite{SekharChivukula:2005xm}. The solid circles represent SU(2) gauge groups, while the dashed circle represents a U(1) gauge group. The lines connecting two circles represent link fields transforming under the adjacent gauge groups. All gauge couplings and $f$-constants are arbitrary parameters.}\label{fig:generalmoose}

The matter field content consists of chiral fermions coupled to the end sites, and vector-like fermions coupled to the internal sites. We adopt the pictorial representation used in Ref.~\cite{SekharChivukula:2006cg}, and shown in Fig.~2, where a lower (upper) line corresponds to a left-handed (right-handed) fermion, and a diagonal dashed line corresponds to a Yukawa coupling. For one generation of quarks or leptons, the corresponding Lagrangian for the mass and Yukawa terms is
\begin{eqnarray}
{\cal L}_{\rm fermion}&=&-\sum_{j=1}^N m_j \bar{\psi'}_{jL}\psi'_{jR}
-\sum_{j=0}^{N-1}f_{j+1}\ y_{j+1}\ \bar{\psi'}_{jL}U_{j+1}\psi'_{(j+1)R} \nonumber \\
&+&f_{N+1} \ \bar{\psi'}_{NL}U_{N+1}\left(\begin{array}{cc} y^u_{N+1} & 0 \\ 0 & y^d_{N+1} \end{array}\right)\psi'_{(N+1)R}+ {\textrm h.c.}\ ,
\label{eq:yukawa}
\end{eqnarray}
where the Yukawa term involving the $U_{j+1}$ link has been appropriately written with a factor of the corresponding $f_{j+1}$ constant. Notice that we only include ``forward'' Yukawa couplings, that is couplings of the type $\bar{\psi'}_{jL}U_{j+1}\psi'_{(j+1)R}$, and not ``backward'' Yukawa couplings, that is couplings of the type $\bar{\psi'}_{jR}U_{j+1}\psi'_{(j+1)L}$. This choice prevents fermion doubling in the mass spectrum ~\cite{Hill:2002me}.

In Eq.~(\ref{eq:yukawa}) $\psi'_{jL}$ and $\psi'_{jR}$ are SU(2) doublets,
\begin{eqnarray}
\psi'_{jL}=\left( \begin{array}{c}
u'_{jL} \\
d'_{jL}
\end{array} \right), \ 
\psi'_{jR}=\left( \begin{array}{c}
u'_{jR} \\
d'_{jR}
\end{array} \right)\ , \nonumber
\label{eq:doublets}
\end{eqnarray}
with the only exception of $\psi'_{(N+1)R}$, which should be interpreted as two SU(2) singlets written in a two-component notation. Notice that in order to obtain the appropriate hypercharge interactions for the light fermions, all doublets must be charged under the U(1) gauge group at the end, with U(1) charge given by the standard model hypercharge of the corresponding left-handed fermion. The U(1) charges of the right-handed singlets, $u'_{(N+1)R}$ and $d'_{(N+1)R}$, are as in the standard model. 

\EPSFIGURE[!t]{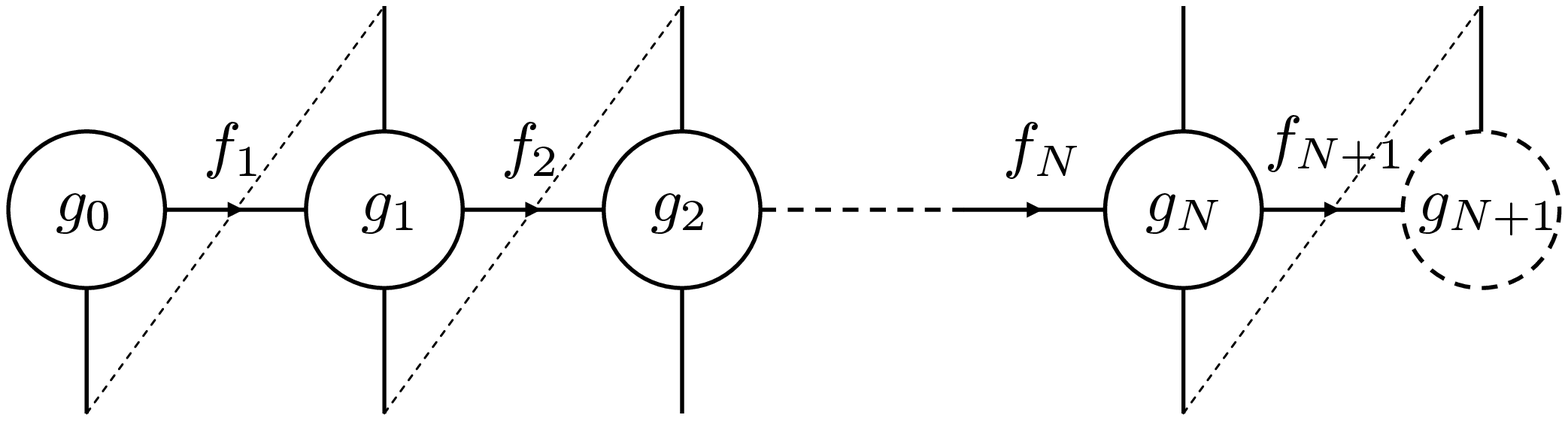,width=14cm,height=4cm}{Moose diagram notation for the coupling of fermion fields to the model of Fig.~1. The lower (upper) segments represent left-handed (right-handed) fermions, while the diagonal dashed lines represent Yukawa couplings of the corresponding fermions to the intersected link field. }\label{fig:fermionmoose}

The Lagrangian of Eq.~(\ref{eq:yukawa}) contains mass terms for the fermions. Diagonalizing the mass-squared matrix gives expressions for the gauge eigenstates in terms of the $N+1$ mass eigenstates,
\begin{eqnarray}
\chi'_{jL} &=& \sum_{n=0}^N \alpha^\chi_{jn}\chi_{nL} \nonumber \\
\chi'_{jR} &=& \sum_{n=0}^N \beta^\chi_{jn}\chi_{nR} \ ,
\label{eq:fermion-expansion}
\end{eqnarray}
where $\chi$ is either $u$ or $d$. Here $\chi_{nL}$ and $\chi_{nR}$ are the left-handed and right-handed components, respectively, of a Dirac fermion, $\chi_n=\chi_{nL}+\chi_{nR}$. We therefore see that, for a given flavor $\chi$, the spectrum consists of $N+1$ Dirac fermions, where the lightest mode, $\chi_0$, is a standard model fermion.\footnote{It is also possible to implement Majorana neutrinos in this scenario, by adding a Majorana mass term at the U(1) site.} We can set the mass of the lightest doublet equal to zero, which in ${\cal L}_{\rm Yukawa}$ can be obtained by setting $y^u_{N+1}=y^d_{N+1}=0$. In this case the coefficients in the expansions preserve the SU(2) structure, $\alpha^u_{jn}=\alpha^d_{jn}\equiv\alpha_{jn}$, $\beta^u_{jn}=\beta^d_{jn}\equiv\beta_{jn}$, and the right-handed components of the lightest fermions are entirely localized on the $(N+1)$-th site. Therefore, using the normalization condition $\sum_{j=1}^{N+1} \beta_{j0}^2=1$, we have  $\beta_{10}=\beta_{20}=\cdots=\beta_{N0}=0$, $\beta_{(N+1)0}=1$, for the lightest doublet.

With $y^u_{N+1}$ and $y^d_{N+1}$ set to zero, the right-handed lightest fermions have no weak isospin charge. Therefore, in this limit the lightest fermion interactions are described by the Lagrangian
\begin{eqnarray}
{\cal L}_{EW}=J^\mu_a\left(\sum_{j=0}^N \alpha_{j0}^2 W^a_{j\mu}\right)+J^\mu_Y W^3_{(N+1)\mu}\ ,
\label{eq:isotriplet}
\end{eqnarray}
where $J^\mu_a$ and $J^\mu_Y$ are the usual isospin and hypercharge currents, respectively, and the normalization condition $\sum_{j=0}^N \alpha_{j0}^2=1$ has been used. At tree-level, four fermion processes are described by the neutral current effective Lagrangian
\begin{eqnarray}
{\cal L}_{NC}&=&-{1\over 2}\left[\sum_{i,j=0}^N \alpha_{i0}^2 \alpha_{j0}^2 <W^3_i W^3_j>\right]J^\mu_3 J_{3\mu}
-\left[\sum_{j=0}^N \alpha_{j0}^2 <W^3_j W^3_{N+1}>\right]J^\mu_3 J_{Y\mu} \nonumber \\
&-&{1\over 2}\left[<W^3_{N+1} W^3_{N+1}>\right]J^\mu_Y J_{Y\mu} \ ,
\label{eq:eff-neutral}
\end{eqnarray}
and the charged current effective Lagrangian
\begin{eqnarray}
{\cal L}_{CC}&=&-{1\over 2}\left[\sum_{i,j=0}^N \alpha_{i0}^2\alpha_{j0}^2 <W^+_i W^-_j>\right]J^\mu_+ J_{-\mu} \ ,
\label{eq:eff-charged}
\end{eqnarray}
where $<W^a_i W^b_j>$ denotes the coefficient of $-i g^{\mu\nu}$ in the $W^{a\mu}_i W^{b\nu}_j$ correlation function\footnote{The $q^\mu q^\nu$ term gives negligible contributions, for external light fermions.}. This Lagrangian is also valid at one-loop order if we neglect vertex and box corrections, which is a good approximation if the loops involve new-physics heavy particles~\cite{Kennedy:1988sn}. In terms of weak and electromagnetic currents, the neutral current Lagrangian of Eq.~(\ref{eq:eff-neutral}) is
\begin{eqnarray}
& & {\cal L}_{NC}= \nonumber \\
& & -{1\over 2}\left[\sum_{i,j=0}^N \alpha_{i0}^2 \alpha_{j0}^2 <W^3_i W^3_j>-2\sum_{j=0}^N \alpha_{j0}^2 <W^3_j W^3_{N+1}>
+<W^3_{N+1} W^3_{N+1}>\right]J^\mu_3 J_{3\mu} \nonumber \\
& & -\left[\sum_{j=0}^N \alpha_{j0}^2 <W^3_j W^3_{N+1}>-<W^3_{N+1} W^3_{N+1}>\right]J^\mu_3 J_{Q\mu}
-{1\over 2}\left[<W^3_{N+1} W^3_{N+1}>\right]J^\mu_Q J_{Q\mu} \nonumber \\
\label{eq:eff-neutral-3Q}
\end{eqnarray}

The $\rho$ parameter is the ratio of the isotriplet neutral current and charged current interactions at zero momentum:
\begin{eqnarray}
\rho=\lim_{q^2\rightarrow 0}{\sum_{i,j=0}^N \alpha_{i0}^2 \alpha_{j0}^2 <W^3_i W^3_j>-2\sum_{j=0}^N \alpha_{j0}^2 <W^3_j W^3_{N+1}>
+<W^3_{N+1} W^3_{N+1}>\over \sum_{i,j=0}^N \alpha_{i0}^2\alpha_{j0}^2 <W^+_i W^-_j>} \ . \nonumber \\
\label{eq:rho}
\end{eqnarray}
It has been proved in Ref.~\cite{SekharChivukula:2005xm} that this quantity is exactly equal to one at tree-level, for arbitrary values of the model parameters. This is a consequence of the approximate SU(2) custodial symmetry of the model, which becomes exact when the hypercharge, and the Yukawa interactions involving the U(1) site are turned off. Moreover, with an appropriate fermion delocalization, that is, with an appropriate choice of the coefficients $\alpha_{j0}$, three of the four leading zero-momentum parameters defined by Barbieri {\em et.al.}~\cite{Barbieri:2004qk} vanish. This occurs when the left-handed light fermion profile resembles the profile of the electroweak bosons, because in such case the light fermions become orthogonal to the heavy vector bosons, and therefore decouple almost entirely from the new physics.

\section{One-Loop Corrections to $\rho$}\label{sec:one-loop}
Having established that the $\rho$ parameter is exactly one at tree-level, it is now interesting to compute one-loop corrections. We do this in a flat Higgsless model, with large ``brane kinetic terms''. By this we mean that all internal gauge couplings, $f$-constants, Dirac masses, and Yukawa couplings do not depend on the site index $j$, while the gauge couplings of $W'^a_0$ and $W'^3_{N+1}$, together with the Yukawa couplings connecting  
$\psi'_{0L}$ with $\psi'_{1R}$, and $\psi'_{NL}$ with $\psi'_{(N+1)R}$ are chosen to be smaller than the corresponding internal parameters. The gauge-sector Lagrangian is
\begin{eqnarray}
{\cal L}_{\rm gauge}=&-&{1\over 4 g^2}W'^a_{0\mu\nu}W'^{a\mu\nu}_0
-{1\over 4\tilde{g}^2} \sum_{j=1}^{N}W'^a_{j\mu\nu}W'^{a\mu\nu}_j
-{1\over 4g^{\prime 2}}W'^3_{(N+1)\mu\nu}W'^{3\mu\nu}_{N+1} \nonumber \\
&+&{f^2\over 4}\sum_{j=1}^{N+1}{\textrm tr}\left(D_\mu U_j\right)^\dagger D^\mu U_j\ ,
\label{eq:gauge-lagrangian-flat}
\end{eqnarray}
where $g^2,g^{\prime 2}\ll\tilde{g}^2/(N+1)$. The coefficients $a_{jn}$ and $b_{jn}$ of Eq.~(\ref{eq:gauge-expansion}) can be calculated perturbatively in $x^2\equiv g^2/\tilde{g}^2$. It is clear that the SU(2) and U(1) gauge groups at the chain ends act approximately as the standard model SU(2)$_{\textrm L}\times$U(1) gauge group, and the internal SU(2) groups act approximately as the new physics. Then the numerical values of $g$ and $g^{\prime}$ will be close to the corresponding standard model values~\cite{Foadi:2003xa}.  

The fermion-sector Lagrangian, for the mass and Yukawa terms, is
\begin{eqnarray}
{\cal L}_{\rm fermion}&=&M\Bigg[\varepsilon_L \bar{\psi'}_{0L}U_1\psi'_{1R}
+\sum_{j=1}^N \bar{\psi'}_{jL}\psi'_{jR}+\sum_{j=1}^{N-1}
\bar{\psi'}_{jL}U_{j+1}\psi'_{(j+1)R} \nonumber \\
&+&\bar{\psi'}_{NL}U_{N+1}
\left(\begin{array}{cc}\varepsilon_{u_R} & 0 \\ 0 & \varepsilon_{d_R}\end{array}\right)\psi'_{(N+1)R}+{\textrm h.c.}\Bigg] \ ,
\label{eq:yukawa-flat}
\end{eqnarray}
where $\varepsilon_L^2 ,\varepsilon_{\chi_R}^2 \ll 1/(N+1)$. The coefficients $\alpha^\chi_{jn}$ and $\beta^\chi_{jn}$ of Eq.~(\ref{eq:fermion-expansion}) can be calculated perturbatively in these parameters. With this set up it is clear that the fermions localized at the chain ends act approximately as the standard model fermions, while the fermions coupled to the internal SU(2) groups are mainly superpositions of the new heavy fermions. Notice that for $\varepsilon_L=0$, both $u_0$ and $d_0$, in the expansions of Eq.~(\ref{eq:fermion-expansion}), are massless, since the corresponding mass matrices have zero determinant. For $\varepsilon_{\chi_R}=0$,only  $\chi_0$ is massless. Then we expect 
\begin{eqnarray}
m_{\chi_0}\propto M \varepsilon_L\varepsilon_{\chi_R} \ ,
\label{eq:lightest}
\end{eqnarray}
to leading order in $\varepsilon_L^2$ and $\varepsilon_{\chi_R}^2$. It is therefore a different value of $\varepsilon_{\chi_R}$ within an SU(2) doublet, $\varepsilon_{u_R}\neq\varepsilon_{d_R}$, which encodes the violation of weak isospin. Moreover, to the extent that we can neglect $m_{\chi_0}$, the corresponding value of $\varepsilon_{\chi_R}$ can be neglected as well. It is then clear that the standard model and new physics contributions to the $\rho$ parameter are mainly due to loops with top modes, $t_k$, and bottom modes, $b_k$, where $k=0,1,2,...N$, and $t_0$, $b_0$ are the standard model top and bottom quarks, respectively. In fact, for a $(u,d)$ doublet we expect $\rho=1$ for the unbroken isospin limit, $\varepsilon_{u_R}=\varepsilon_{d_R}$, and for light fermions we have just argued that $\varepsilon_{u_R}\simeq\varepsilon_{d_R}\simeq 0$. 

To leading order we can assume that the light left-handed fermions are exactly localized at the $j=0$ site, $a_{j0}\rightarrow\delta_{j0}$. Then Eq.~(\ref{eq:rho}) becomes
\begin{eqnarray}
\rho=\lim_{q^2\rightarrow 0}{<W^3_0 W^3_0>-2<W^3_0 W^3_{N+1}>+<W^3_{N+1} W^3_{N+1}>\over <W^+_0 W^-_0>} \ . \nonumber \\
\label{eq:rho-eL=0}
\end{eqnarray}
To leading order we can also take $x\rightarrow 0$, in which case
\begin{eqnarray}
W^3_0&=&e A + {g^2\over\sqrt{g^2+g^{\prime 2}}} Z \nonumber \\
W^3_{N+1}&=&e A -{g^{\prime 2}\over\sqrt{g^2+g^{\prime 2}}} Z \nonumber \\
W^\pm_0&=&g W \nonumber \\
{m_W^2\over m_Z^2}&=&{g^2\over g^2+g^{\prime 2}} \ ,
\label{eq:settings}
\end{eqnarray}
where $A$, $Z$, and $W$ are the ordinary electroweak bosons. Inserting these expressions in Eq.~(\ref{eq:rho-eL=0}), we see that the photon contribution vanishes, as it must. Then, expanding the  $W$ and $Z$ propagators, we obtain, for $\Delta\rho\equiv\rho-1$,
\begin{eqnarray}
\Delta\rho={\Pi_{WW}(0)\over m_W^2}-{\Pi_{ZZ}(0)\over m_Z^2} \ ,
\label{eq:one-loop-rho}
\end{eqnarray}
where $\Pi_{WW}$ and $\Pi_{ZZ}$ are the coefficients of $i g_{\mu\nu}$ in the 1PI $W$ and $Z$ functions, respectively. Notice that this equation for $\Delta\rho$ contains loops in the $W$ and $Z$ boson propagators only, and thus corresponds to the Peskin-Takeuchi $\alpha T$ parameter. In fact we are considering the leading order term in an expansion in $x^2$, which amounts to ignoring the small contribution from the heavy boson propagators. Including higher modes in the expansions of Eq.~(\ref{eq:settings}), and considering that the coupling of the heavy vector bosons to the heavy fermions is of order $\tilde{g}$, it can be shown that the heavy modes give corrections of order ${\cal O}(x^4)$ to Eq.~(\ref{eq:one-loop-rho}). Therefore, to order ${\cal O}(x^2)$ we have $\Delta\rho=\alpha T$ in this model~\cite{Chivukula:2004af}~\cite{SekharChivukula:2004mu}~\cite{SekharChivukula:2005xm}. 

\EPSFIGURE[!t]{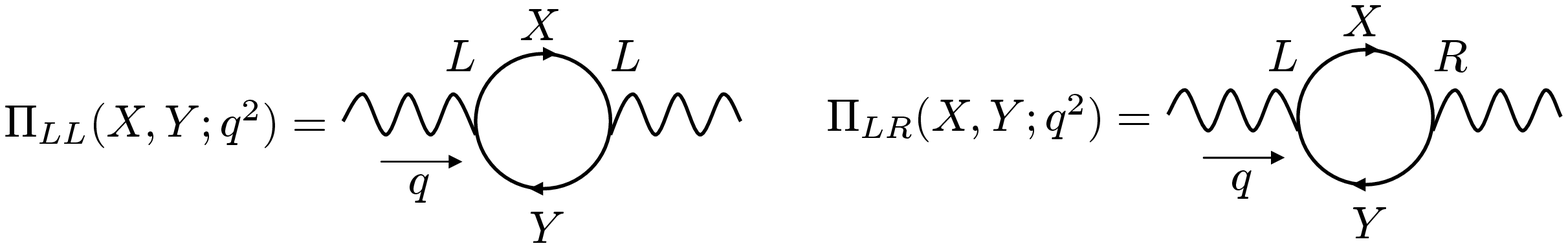,width=14cm,height=2.5cm}{Vacuum polarization amplitudes for left-left and left-right gauge currents.}\label{fig:loops}

Notice that since we take $q^2=0$, only the isospin part contributes in $\Pi_{ZZ}$. We define $\Pi_{LL}(X,Y;q^2)$ as the coefficient of $i g_{\mu\nu}$ in the vacuum polarization amplitude with left-handed currents only, and fermions $X$ and $Y$ in the loop. In a similar way we define $\Pi_{LR}$, as shown in Fig.~3, while it can be easily proved that $\Pi_{RR}=\Pi_{LL}$. (Trivially, $\Pi_{RL}=\Pi_{LR}$.) At zero momentum these functions are~\cite{Peskin:1995ev}
\begin{eqnarray}
\Pi_{LL}(0)&=&{1\over 16\pi^2}\left[(m_X^2+m_Y^2)E-2\left(m_X^2 b_1(m_X,m_Y;0) + m_Y^2 b_1(m_Y,m_X;0)\right)\right] \nonumber \\
\Pi_{LR}(0)&=&{1\over 16\pi^2}\left[-2 m_X m_Y E + 2 m_X m_Y b_0(m_X,m_Y;0)\right] \ ,
\label{eq:VPA}
\end{eqnarray}
where
\begin{eqnarray}
b_0(m_X,m_Y;q^2)&=&\int_0^1 dx \log\left({x\ m_X^2 + (1-x) m_Y^2 - x(1-x)q^2\over\mu^2}\right) \nonumber \\
b_1(m_X,m_Y;q^2)&=&\int_0^1 dx\ x \log\left({x\ m_X^2 + (1-x) m_Y^2 - x(1-x)q^2\over\mu^2}\right) \ .
\label{eq:b}
\end{eqnarray}
Here $E$ is the divergent part of the loop diagram from dimensional regularization , $E=\frac{2}{\epsilon}-\gamma+\log(4\pi)-\log(\mu^{2})$ ($\epsilon=4-d)$, and $\mu$ is the renormalization mass scale. The couplings constants are formally given by
\begin{eqnarray}
g^{CC}_{L(u_k,d_l)}=\sum_{j=0}^N \alpha^u_{jk} \alpha^d_{jl} a_{j0} \quad , \quad
g^{CC}_{R(u_k,d_l)}=\sum_{j=1}^N \beta^u_{jk} \beta^d_{jl} a_{j0} \ ,
\label{eq:W-couplings}
\end{eqnarray}
for the left-handed and right-handed couplings of $u_k$ and $d_l$ to the $W$ boson, and
\begin{eqnarray}
g^{NC}_{L(\chi_k,\chi_l)}=\sum_{j=0}^N \alpha^\chi_{jk} \alpha^\chi_{jl} (b_{j0}-b_{(N+1)0}) \quad , \quad
g^{NC}_{R(\chi_k,\chi_l)}=\sum_{j=1}^N \beta^\chi_{jk} \beta^\chi_{jl} (b_{j0}-b_{(N+1)0}) \ ,
\label{eq:Z-couplings}
\end{eqnarray}
for the left-handed and right-handed couplings of $\chi_k$ and $\chi_l$ to the $Z$ boson. These expressions can be used to find the couplings perturbatively in the small parameters. Once this is done, the 1PI functions can be computed by
\begin{eqnarray}
& & \Pi_{WW}(0)= \nonumber \\
& & \sum_{k,l} {3\over 2}\Bigg[\left(\left(g^{CC}_{L(t_k,b_l)}\right)^2+\left(g^{CC}_{R(t_k,b_l)}\right)^2\right)
\Pi_{LL}(t_k,b_l;0)
+ 2 g^{CC}_{L(t_k,b_l)} g^{CC}_{R(t_k,b_l)} \Pi_{LR}(t_k,b_l;0) \Bigg] \  \nonumber \\
\label{eq:one-loop-WW}
\end{eqnarray} 
for the $W$ boson, and
\begin{eqnarray}
& & \Pi_{ZZ}(0)= \nonumber \\
& & \sum_{k,l} {3\over 4}\Bigg[\left(\left(g^{NC}_{L(t_k,t_l)}\right)^2+\left(g^{NC}_{R(t_k,t_l)}\right)^2\right)
\Pi_{LL}(t_k,t_l;0)
+ 2 g^{NC}_{L(t_k,t_l)} g^{NC}_{R(t_k,t_l)} \Pi_{LR}(t_k,t_l;0) \nonumber \\
& & + \left(\left(g^{NC}_{L(b_k,b_l)}\right)^2+\left(g^{NC}_{R(b_k,b_l)}\right)^2\right)\Pi_{LL}(b_k,b_l;0) 
+ 2 g^{NC}_{L(b_k,b_l)} g^{NC}_{R(b_k,b_l)} \Pi_{LR}(b_k,b_l;0) \Bigg] \ .
\label{eq:one-loop-ZZ}
\end{eqnarray} 
for the $Z$ boson. In these expressions the factor 3 takes into account the different color contributions, the factor 1/2 in $\Pi_{WW}$ comes from 1/$\sqrt{2}$ in the Lagrangian, and the factor 1/4 in $\Pi_{ZZ}$ is the product of isospin quantum numbers.

Inserting Eqs.~(\ref{eq:one-loop-WW}) and (\ref{eq:one-loop-ZZ}) in Eq.~(\ref{eq:one-loop-rho}), and using Eq.~(\ref{eq:VPA}), gives the following expression for the infinite part of $\Delta\rho$:
\begin{eqnarray}
\sum_{k,l}&&\left[{\left(g^{CC}_{L(t_k,b_l)}\right)^2+\left(g^{CC}_{R(t_k,b_l)}\right)^2\over m_W^2}{m_{t_k}^2+m_{b_l}^2\over 2}
-{g^{CC}_{L(t_k,b_l)} g^{CC}_{R(t_k,b_l)}\over m_W^2} m_{t_k} m_{b_l} \right. \nonumber \\
&&-{\left(g^{NC}_{L(t_k,t_l)}\right)^2+\left(g^{NC}_{R(t_k,t_l)}\right)^2\over m_Z^2}{m_{t_k}^2+m_{t_l}^2\over 4}
+{g^{NC}_{L(t_k,t_l)} g^{NC}_{R(t_k,t_l)}\over m_Z^2} {m_{t_k} m_{t_l}\over 2} \nonumber \\
&& \left.  -{\left(g^{NC}_{L(b_k,b_l)}\right)^2+\left(g^{NC}_{R(b_k,b_l)}\right)^2\over m_Z^2}{m_{b_k}^2+m_{b_l}^2\over 4}
+{g^{NC}_{L(b_k,b_l)} g^{NC}_{R(b_k,b_l)}\over m_Z^2} {m_{b_k} m_{b_l}\over 2} \right] \ .
\label{eq:infinite}
\end{eqnarray}
This can be proved to be exactly zero, for $x\rightarrow 0$, to all orders in $\varepsilon_L$, $\varepsilon_{t_R}$, and $\varepsilon_{b_R}$, by using recurrence and completeness relations for the expansion coefficients of the top and bottom towers.\footnote{This was independently proved, in private communications, by R. S. Chivukula for the deconstructed model, and by one of the authors of this note (R. Foadi) for the continuum model.} Notice that even though we approximated the left-handed light fermions to be exactly localized, which corresponds to setting $\varepsilon_L=0$, we could have taken different $\varepsilon_L$'s for the light fermions and for the third-generation quarks, and set only the former equal to zero. Then the $\Delta\rho$ we compute here would be valid at all orders in the top-bottom $\varepsilon_L$.

The next step is to calculate the finite part of $\Delta\rho$. We do it analytically for $N=1$ and $N\rightarrow\infty$, and show numerical results for arbitrary values of $N$. Since we are only interested in the leading order new-physics contribution, we take $\varepsilon_L\rightarrow 0$, since a finite $\varepsilon_L$ would give, to leading order, the ordinary standard model contribution. To see why, we first neglect the bottom mass, which in our language amounts to setting $\varepsilon_{b_R}=0$. Then the standard model contribution to $\Delta\rho$ is proportional to $m_t^2$  and thus, by Eq.~(\ref{eq:lightest}), to $M^2\varepsilon_L^2\varepsilon_{t_R}^2$. On the other hand, in the $\varepsilon_L\rightarrow 0$ limit the heavy top and bottom modes are not degenerate (since $m_{t_k}\neq m_{b_k}$ for $\varepsilon_{t_R}\neq\varepsilon_{b_R}$), and give a non-zero contribution to $\Delta\rho$. The latter is therefore the leading order new-physics contribution~\cite{SekharChivukula:2006cg}.

\subsection{$N=1$}
In the three-site model the gauge sector consists of the ordinary electroweak gauge bosons, a heavy charged $W_1$ boson, and a heavy neutral $Z_1$ boson\footnote{The content of this section reproduces the analysis of Ref.~\cite{SekharChivukula:2006cg}.}. In the expansions of Eq.~(\ref{eq:gauge-expansion}), we only need the coefficients of the $W$ and the $Z$ bosons, $a_{j0}$ and $b_{j0}$. In the $x\rightarrow 0$ limit these are~\cite{Foadi:2003xa}
\begin{eqnarray}
a_{00} = g \quad , \quad a_{10} = {g\over 2}
\label{eq:gauge-expansion-3-charged}
\end{eqnarray}
for the $W$ boson, and
\begin{eqnarray}
b_{00} = {g^2\over\sqrt{g^2+g^{\prime 2}}} \quad , \quad
b_{10} = {1\over 2} {g^2-g^{\prime 2}\over \sqrt{g^2+g^{\prime 2}}}  \quad , \quad
b_{20} = - {g^{\prime 2}\over\sqrt{g^2+g^{\prime 2}}}
\label{eq:gauge-expansion-3-neutral}
\end{eqnarray}
for the $Z$ boson. The $W$ and $Z$ masses are
\begin{eqnarray}
m_W^2 = g^2{f^2\over 8} \quad , \quad
m_Z^2 = (g^2+g^{\prime 2}){f^2\over 8} \ .
\label{eq:gauge-masses-3}
\end{eqnarray}

For a given fermion flavor $\chi$, the spectrum consists of two Dirac fermions: a light state, $\chi_0=\chi_{0L}+\chi_{0R}$, which will be identified with a standard model fermion, and a heavy state, $\chi_1=\chi_{1L}+\chi_{1R}$. In the $\varepsilon_L\rightarrow 0$ limit the coefficients of Eq.~(\ref{eq:fermion-expansion}) are~\cite{SekharChivukula:2006cg}
\begin{eqnarray}
& & \alpha^\chi_{00} = -1 \quad , \quad
\alpha^\chi_{10} = 0 \nonumber \\
& & \alpha^\chi_{01} = 0 \quad , \quad
\alpha^\chi_{11} = -1
\label{eq:left-fermion-expansion-3}
\end{eqnarray}
for the left-handed components, and
\begin{eqnarray}
& & \beta^\chi_{10}= -{\varepsilon_{\chi_R}\over\sqrt{1+\varepsilon_{\chi_R}^2}} \quad , \quad
\beta^\chi_{20} = {1\over\sqrt{1+\varepsilon_{\chi_R}^2}} \nonumber \\
& & \beta^\chi_{11} = {1\over\sqrt{1+\varepsilon_{\chi_R}^2}} \quad , \quad
\beta^\chi_{21} = -{\varepsilon_{\chi_R}\over\sqrt{1+\varepsilon_{\chi_R}^2}}
\label{eq:right-fermion-expansion-3}
\end{eqnarray}
for the right handed components. The $\chi_0$ and $\chi_1$ masses are
\begin{eqnarray}
m_{\chi_0} = 0 \quad , \quad
m_{\chi_1} = M \sqrt{1+\varepsilon_{\chi_R}^2} \ .
\label{eq:fermion-masses-3}
\end{eqnarray}
Notice that with $\varepsilon_L$ set to zero, the mass of the lightest mode is zero, top quark included. This is fine, since we are looking for the new physics contribution to $\Delta\rho$, which is due to the heavy modes.

Inserting these results in Eq.~(\ref{eq:W-couplings}) gives
\begin{eqnarray}
\begin{array}{ll}
g^{CC}_{L(t_0,b_0)}=g & 
g^{CC}_{R(t_0,b_0)}=0 \\
g^{CC}_{L(t_0,b_1)}=0 & 
g^{CC}_{R(t_0,b_1)}=-g\left(\varepsilon_{t_R}/ 2\sqrt{1+\varepsilon_{t_R}^2}\right) \\
g^{CC}_{L(t_1,b_0)}=0 & 
g^{CC}_{R(t_1,b_0)}=0 \\
g^{CC}_{L(t_1,b_1)}=g/2 & 
g^{CC}_{R(t_1,b_1)}=g\left(1/ 2\sqrt{1+\varepsilon_{t_R}^2}\right)
\end{array}
\label{eq:W-couplings-3}
\end{eqnarray}for the couplings to the $W$ boson, and
\begin{eqnarray}
\begin{array}{ll}
g^{NC}_{L(t_0,t_0)}=\sqrt{g^2+g^{\prime 2}} & 
g^{NC}_{R(t_0,t_0)}=\sqrt{g^2+g^{\prime 2}}\left(\varepsilon_{t_R}^2/ 2(1+\varepsilon_{t_R}^2)\right) \\
g^{NC}_{L(t_0,t_1)}=0 & 
g^{NC}_{R(t_0,t_1)}=-\sqrt{g^2+g^{\prime 2}}\left(\varepsilon_{t_R}/ 2(1+\varepsilon_{t_R}^2)\right) \\
g^{NC}_{L(t_1,t_1)}=\sqrt{g^2+g^{\prime 2}}/2 & 
g^{NC}_{R(t_1,t_1)}=\sqrt{g^2+g^{\prime 2}}\left(1/2(1+\varepsilon_{t_R}^2)\right) \\
g^{NC}_{L(b_0,b_0)}=\sqrt{g^2+g^{\prime 2}} & 
g^{NC}_{R(b_0,b_0)}=0 \\
g^{NC}_{L(b_0,b_1)}=0 & 
g^{NC}_{R(b_0,b_1)}=0 \\
g^{NC}_{L(b_1,b_1)}=\sqrt{g^2+g^{\prime 2}}/2 & 
g^{NC}_{R(b_1,b_1)}=\sqrt{g^2+g^{\prime 2}}/2
\end{array}
\label{eq:Z-couplings-3}
\end{eqnarray}
for the couplings to the $Z$ boson. Notice that dividing the couplings by the gauge boson masses, as demanded by Eq.~(\ref{eq:one-loop-rho}), completely removes $g$ and $g^{\prime}$.

Using Eqs.~(\ref{eq:fermion-masses-3}) - (\ref{eq:Z-couplings-3}), it can be shown that Eq.~(\ref{eq:infinite}) is indeed satisfied, and the infinite part is canceled out. The finite part may be obtained by inserting the gauge couplings into the expressions for $\Pi_{WW}(0)$ and $\Pi_{ZZ}(0)$, Eq.~(\ref{eq:one-loop-WW}) and Eq.~(\ref{eq:one-loop-ZZ}), respectively, and these into Eq.~(\ref{eq:one-loop-rho}). Expanding in $\varepsilon_{t_R}$, the new-physics leading contribution to $\Delta\rho$, for $N=1$, is found to be
\begin{eqnarray}
\Delta\rho(1)={1\over 16\pi^2} {\varepsilon_{t_R}^4 M^2 \over v^2} \ ,
\label{eq:delta-rho-3}
\end{eqnarray}
where $v$ is the ordinary standard model vacuum expectation value, $v=\textrm{246 GeV}$, and is related to $f$ in the $N$-site model by $v^2=f^2/(N+1)$.

\subsection{Arbitrary $N$}

\EPSFIGURE[!t]{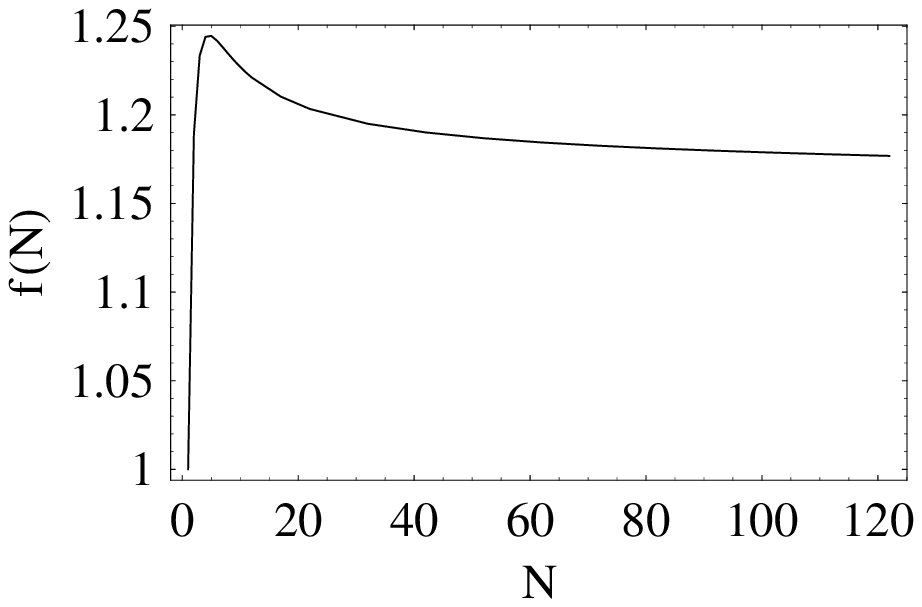,width=10cm,height=6cm}{Plot of $f(N)$, defined by Eq.~(\ref{eq:NDr}). The continuum limit gives approximately a correction of 17\%, with respect to the three-site model.}\label{fig:variation}

The $N$-site model, for arbitrary values of $N$, involves complicated expressions, especially in the fermion sector. We therefore opt for a numerical computation of $\Delta\rho$. As for the three-site model ($N$=1), we do this in the limit $\varepsilon _L, x  \to 0$, since this gives the leading order new-physics contribution. Based on general arguments~\cite{SekharChivukula:2006cg}, and on the results of the last section, we expect a result of the form
\begin{equation}
\Delta \rho \left( N \right) = {f( N)\over 16\pi^2}{\varepsilon _{t_R}^4 M^2 \over v^2} \ ,
\label{eq:NDr}
\end{equation}
to leading order in $\varepsilon _{t_R}$, where $f(N)$ is the quantity we set out to find. Since in a numerical calculation are included not only the leading term but also higher order corrections, we need $\varepsilon_L^2,\ x^2,$ and $\varepsilon_{t_R}^2$ to be much smaller than one, in order to make the non-leading order contributions negligible, and recover the analytical results for the three-site model and the continuum model. Also, since we work in the limit $\varepsilon _L ,\ x  \to 0$, we take $\varepsilon _L$ and $x$ much smaller than $\varepsilon_{t_R}$. Therefore, we arbitrarily choose the values $x,\varepsilon_L\sim 10^{-5}$, $\varepsilon_{t_R}\sim 10^{-3}$, and calculate $\Delta\rho$. Dividing the result by $\epsilon_{tR}^4 M^2/16\pi^2 v^2$ gives $f(N)$. To get an estimate of the error of this, we performed also a semi-analytical calculation of $f(N)$, for $N=3,4,...,10$, by approximating irrational numbers with rationals having fifty significant figures. The error on $f(N)$ for the numerical calculation compared to the semi-analytical one was approximately constant and equal to $0.1\%$. We show our results in Fig.~4, for $N$ between 1 and 122.\footnote{The reason to stop at $N=122$ is simply that the percentage increase in time to complete the calculation was becoming much larger than the very small percentage change in $f\left( N\right)$.} For the three-site model we obtain $f(1)=1$, in agreement with the analytical result found in the last section. Moreover we find $f(122)=1.177$, which is very close to the value we will find for the continuum model ($N\rightarrow\infty$) in the next section.

\subsection{$N\rightarrow\infty$}
The Lagrangian for the gauge sector, Eq.~(\ref{eq:gauge-lagrangian-flat}), and the Yukawa Lagrangian, Eq.~(\ref{eq:yukawa-flat}), have a well defined 
$N\rightarrow\infty$ limit, provided that $\tilde{g},f\sim(N+1)^{1/2}$, $M\sim (N+1)$, and $\varepsilon_L,\varepsilon_{\chi_R}\sim (N+1)^{-1/2}$, for large values of $N$~\cite{Foadi:2003xa}. If this is true, the summations are replaced by integrals, and the site index $i$ becomes a continuum variable $y$. The action for the gauge sector becomes
\begin{eqnarray}
{\cal S}_\textrm{gauge}&=& \int d^4x\int_0^{\pi R}dy
\left[-{1\over 4 g_5^2}W^a_{MN}W^{a\,MN}
-\delta(y){1\over 4g^2}W^a_{\mu\nu}W^{a\,\mu\nu}\right.\nonumber\\
&&\qquad\qquad\qquad\qquad\qquad\qquad\qquad\left.-\delta(\pi R-y){1\over 4g'^2}W^3_{\mu\nu}W^{3\,\mu\nu}\right]\ ,
\label{eq:Ninf}
\end{eqnarray}
where $M,N=0,1,2,3,5$, and
\begin{eqnarray}
\pi R=\lim_{N\rightarrow\infty}{2(N+1)\over\tilde{g}f}\ \ ,\ \ 
g_5^2=\lim_{N\rightarrow\infty}{2\tilde{g}\over f}\ .
\label{eq:4d-5d-gauge}
\end{eqnarray}
These equations show that the condition $g^2,g^{\prime 2}\ll \tilde{g}^2/(N+1)$ of the finite-$N$ model translates into $g^2,g^{\prime 2}\ll g_5^2/\pi R$ in the continuum model. The action for the fermion sector (for one generation) becomes
\begin{eqnarray}
& &{\cal S}_\textrm{fermion} = \int_0^{\pi R}dy\int d^4x
\Bigg[{1\over\pi R}\left(\bar{\psi}i\Gamma^{\mu} D_{\mu}\psi
+{\kappa}\left({1\over 2}\bar{\psi}i\Gamma^5 D_5\psi + \ {\rm h.c.}\right)\right)\nonumber\\
&+&\delta(y){1\over t_L^2}\bar{\psi}_L i \gamma^\mu D_\mu \psi_L
+\delta(\pi R-y)\Bigg({1\over t_{u_R}^2}\bar{u}_R i \gamma^\mu D_\mu u_R
+{1\over t_{d_R}^2}\bar{d}_R i \gamma^\mu D_\mu d_R\Bigg)\Bigg]\ ,\nonumber \\
& &
\label{eq:fermionaction3}
\end{eqnarray}
where $\Gamma^M=(\gamma^\mu,-i \gamma^5)$, and
\begin{eqnarray}
{\pi R\over\kappa}=\lim_{N\rightarrow\infty}{N+1\over M} \ \ ,\ \ 
t_L=\lim_{N\rightarrow\infty}\sqrt{N+1}\ \varepsilon_L \ \ , \ \ 
t_{\chi_R}=\lim_{N\rightarrow\infty}\sqrt{N+1}\ \varepsilon_{\chi_R} \ .
\label{eq:4d-5d-fermion}
\end{eqnarray}
Therefore, the condition $\varepsilon_L^2 ,\varepsilon_{\chi_R}^2 \ll 1/(N+1)$ for the finite-$N$ model translates into $t_L^2,t_{\chi_R}^2\ll 1$.

Notice that the appearance of delta functions is due to the fact that the parameters relative to the end sites are different from the ``bulk'' parameters. Notice also that although these actions seem to describe a five-dimensional Yang-Mills theory, this is actually not true for two reasons. First, as in the finite-$N$ model, all left-handed fermions must couple to the gauge field at $y=\pi R$, thereby introducing non-local interactions, from a five-dimensional standpoint. Second, a non-zero value of $\kappa-1$, in Eq.~(\ref{eq:fermionaction3}), parametrizes a local breaking of the five-dimensional Lorentz invariance, in addition to the non-local breaking due to compactification. Therefore, this theory should be interpreted as a model from continuum theory space.

After choosing a gauge in which $W^a_5\equiv 0$, all gauge field four-dimensional components can be expanded in towers of heavy vector bosons, in analogy with the expansions of Eq.~(\ref{eq:gauge-expansion}), with the difference that the towers are now infinite:
\begin{eqnarray}
W^\pm_\mu(x,y)&=&\sum_{n=0}^\infty a_n(y) W^\pm_{n\mu}(x) \nonumber \\
W^3_\mu(x,y)&=&eA_\mu(x)+\sum_{n=0}^\infty b_n(y) Z_{n\mu}(x) \ .
\label{eq:gauge-expansion-inf}
\end{eqnarray}
Similarly, the fermion fields can be expanded in infinite towers of massive Dirac fermions, in analogy with the expansions of Eq.~(\ref{eq:fermion-expansion}):
\begin{eqnarray}
\chi_L(x,y) &=& \sum_{n=0}^\infty \alpha^\chi_n(y)\chi_{nL}(x) \nonumber \\
\chi_R(x,y) &=& \sum_{n=0}^\infty \beta^\chi_n(y)\chi_{nR}(x) \ .
\label{eq:fermion-expansion-inf}
\end{eqnarray}
Eq.~(\ref{eq:one-loop-WW}) and Eq.~(\ref{eq:one-loop-ZZ}) are still valid, as long as the formulas for the coupling constants, Eq.~(\ref{eq:W-couplings}) and Eq.~(\ref{eq:Z-couplings}), are replaced respectively by
\begin{eqnarray}
g^{CC}_{L(u_k,d_l)}&=&\int_0^{\pi R}dy\left[{1\over \pi R}+{\delta(y)\over t_L^2}\right] \alpha^u_k(y) \alpha^d_l(y) a_0(y)\nonumber \\
g^{CC}_{R(u_k,d_l)}&=&\int_0^{\pi R}dy{1\over \pi R} \beta^u_k(y) \beta^d_l(y) a_0(y) \ ,
\label{eq:W-couplings-inf}
\end{eqnarray}
and
\begin{eqnarray}
g^{NC}_{L(\chi_k,\chi_l)}&=&\int_0^{\pi R}dy\left[{1\over \pi R}+{\delta(y)\over t_L^2}\right] 
\alpha^u_k(y) \alpha^d_l(y) \left(b_0(y)-b_0(\pi R)\right)\nonumber \\
g^{NC}_{L(\chi_k,\chi_l)}&=&\int_0^{\pi R}dy{1\over \pi R} \beta^u_k(y) \beta^d_l(y)\left(b_0(y)-b_0(\pi R)\right) \ .
\label{eq:Z-couplings-inf}
\end{eqnarray}
All wavefunctions, and therefore all couplings, can be calculated perturbatively in the small parameters. Once this is done, the computation of $\Pi_{WW}(0)$ and $\Pi_{ZZ}(0)$, and thus of $\Delta\rho$ can be carried out. We find that the infinite part cancels out, in agreement with the general result for arbitrary $N$, while the finite part is\footnote{The numerical factor $\beta$ is just the sum of a complicated numerical series.}
\begin{eqnarray}
\Delta\rho = {\beta\over 16\pi^2} {t_{t_R}^4\over \left(\pi R/\kappa\right)^2 v^2} \ \ ,\ \ \beta=1.1724 \ .
\label{eq:delta-rho-inf}
\end{eqnarray} 
Using Eq.~(\ref{eq:4d-5d-fermion}) we see that this result is in agreement with the result for the $N$-site model, Eq.~(\ref{eq:NDr}), since in the large $N$ limit $f(N)$ indeed approaches $\beta$ (we found $f(122)$=1.177, with the function slowly decreasing to a horizontal asymptote). As we noted in the last section, this result is already very well approximated by the $N=1$ model.

\section{Experimental Bounds on Fermion Masses} \label{sec:experimental}
With these results, the experimental upper bounds on $\Delta\rho$ translate into lower bounds for the mass of the heavy fermions. These can be derived only after a relation between $\varepsilon_L^2$ and $x^2$ is established. In the $N=1$ model such relation is imposed by ideal delocalization, in which case three of the four leading electroweak parameters introduced by Barbieri {\em et.al.}~\cite{Barbieri:2004qk} exactly vanish. In the model with arbitrary $N$, ideal delocalization is not possible, since we have already imposed translational invariance on the "`bulk parameters"'. However we can require that the $S$ parameter vanishes. This is phenomenologically sufficient, since the terms parametrizing low energy four-fermion interactions are naturally suppressed. Taking $\alpha$, $m_Z$, and $m_W$ as fundamental input parameters, a fermion's coupling to the $W$ boson, as a function of the $S$, $T$ and $U$ parameters, is~\cite{Foadi:2004ps}  
\begin{equation}
g_{L0}^{CC}  = \frac{e}
{s}\left[ {1 + \frac{{\alpha S}}
{{4{\kern 1pt} s^2 }} - \frac{{c^2 \alpha T}}
{{2{\kern 1pt} s^2 }} - \frac{{\left( {c^2  - s^2 } \right)\alpha U}}
{{8s^2 }}} \right],
\end{equation}
where $c\equiv m_W/m_Z$ and $s^2\equiv \sqrt{1-c^2}$. Since at tree-level $T,U={\cal O}(x^4)$, we can obtain the leading order expression for $S$ by just computing $g_{L0}^{CC}$. Including corrections of order ${\cal O}(x^2)$ and ${\cal O}(\varepsilon_L^2)$, this is
\begin{equation}
g_{L0}^{CC}  =  g\left[1-{N(2N+1)\over 12(N+1)}x^2-{N\over 2}\varepsilon_L^2\right].
\label{eq:g-expr}
\end{equation}
Expressing $e$ and $s$ in terms of the input parameters (see Ref.~\cite{Foadi:2003xa}), we find that $S$ vanishes if
\begin{equation}
\varepsilon _L^2  = {1\over 3}{N+2\over N+1}\ x^2 \ .
\label{eq:Yahoo-delocal}
\end{equation}
Notice that this gives the correct expressions for the three-site model~\cite{SekharChivukula:2006cg} and the continuum model~\cite{Foadi:2004ps}.
To turn Eq.s~(\ref{eq:g-expr}), (\ref{eq:Yahoo-delocal}) into a bound on $m_f$, we need to gather some additional piece of information. First we need expressions for the $W$ and $W_1$ masses, which can be found in Ref.~\cite{Foadi:2003xa}. To leading order in $x^2$:
\begin{equation}
m_W^2  = \frac{{g^2 f^2 }}
{{4\left( {N + 1} \right)}}\,\,,\,\,\,\,m_{W_1}^2  = \tilde g^2 f^2 \sin ^2\! \left( {\frac{\pi}
{{2\left( {N + 1} \right)}}} \right) \ .
\label{eq:mWWp}
\end{equation}
To leading order in $\varepsilon_L$ and $\varepsilon_{t_R}$ the top mass is
\begin{equation}
m_t = M\ \varepsilon_L\ \varepsilon_{t_R} \ .
\label{eq:top-mass-N}
\end{equation}
The heavy fermions are all approximately degenerate. To leading order in $\varepsilon_L$ and $\varepsilon_{\chi_R}$, the mass $m_{\chi_1}$ of the lightest of these heavy modes is
\begin{equation}
m_{\chi_1}  = 2{\kern 1pt} M{\kern 1pt} \sin\! \left( {\frac{\pi }
{{2{\kern 1pt} \left( {2{\kern 1pt} N + 1} \right)}}} \right).
\label{eq:mtp}
\end{equation}
Recalling that $x\equiv g/\tilde{g}$, and using Eqs.~(\ref{eq:Yahoo-delocal}) - (\ref{eq:mtp}), Eq.~(\ref{eq:NDr}) gives
\begin{equation}
m_{\chi_1}  = \frac{3}
{{8{\kern 1pt} \pi }}{\kern 1pt} \frac{{\sin\! \left( {\frac{\pi }
{{2{\kern 1pt} \left( {2{\kern 1pt} N + 1} \right)}}} \right)}}
{{\left( {N+2} \right){\kern 1pt} \sin ^2\! \left( {\frac{\pi }
{{2{\kern 1pt} \left( {N + 1} \right)}}} \right)}}{\kern 1pt} \sqrt {\frac{{f\left( N \right)}}
{{\Delta \rho }}} {\kern 1pt} \frac{{m_t^2 }}
{{v{\kern 1pt} m_W^2 }} m_{W_1}^2\,.
\label{eq:bound}
\end{equation}

We see explicitly that the upper bounds on $\Delta\rho$ become, for fixed values of $m_{W_1}$ and $N$, lower bounds on $m_{\chi_1}$. The experimental bounds on the $\rho$ parameter depend on the value of the reference Higgs mass, $m_H^\textrm {ref}$. For $m_H^2\gg m_W^2$ the Higgs contribution to $\Delta\rho$ is
\begin{eqnarray}
\left(\Delta\rho\right)_\textrm {Higgs} = -{3\alpha\over 16\pi c^2}\log{m_H^2\over m_W^2} \ .
\label{eq:higgs-rho}
\end{eqnarray}
In our Higgsless model the contribution to $\Delta\rho$ from the $W_1$ boson, for $m_{W_1}^2\gg m_W^2$, has the same form, with exactly the same coefficient~\cite{Matsuzaki:2006wn}:
\begin{eqnarray}
\left(\Delta\rho\right)_{\textrm W_1} = -{3\alpha\over 16\pi c^2}\log{m_{W_1}^2\over m_W^2} \ .
\label{eq:higgs-W1}
\end{eqnarray} 
We therefore interpret the phenomenological bounds on $\Delta\rho$, extracted for a given value of $m_H^\textrm {ref}$, as bounds extracted for the same value of the $W_1$ mass, $m_{W_1}=m_H^\textrm {ref}$. 

\EPSFIGURE[!t]{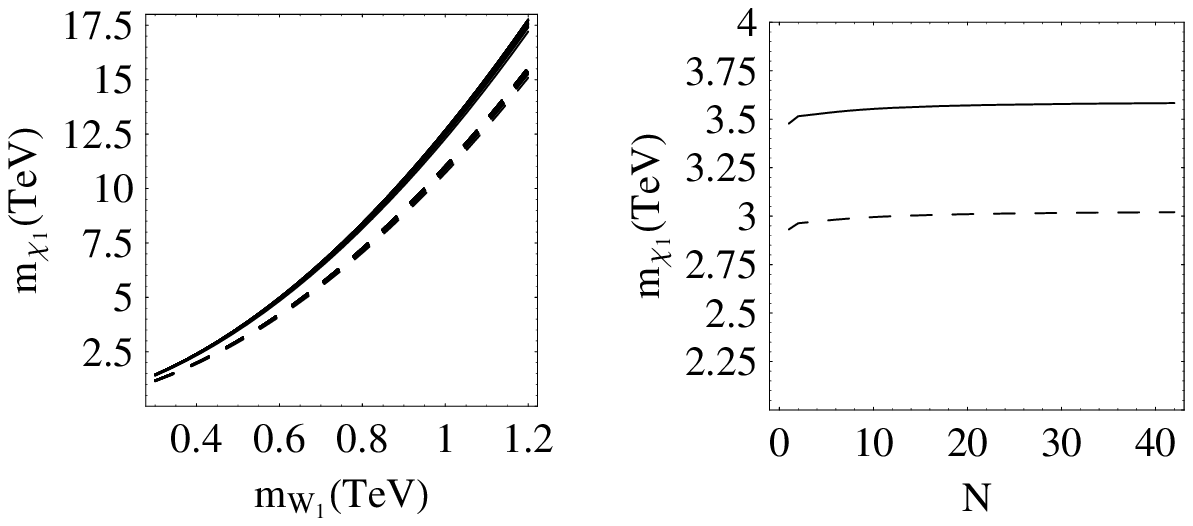,width=15cm,height=7cm}{Lower bounds on the mass $m_{\chi_1}$ of the lightest among the heavy fermions, as a function of $m_{W_1}$, with $N$ varying between 1 and 122 (left). We also plot the same quantity as a function of $N$, for $m_{W_1}$=500 GeV (right). In each case a solid line corresponds to $\Delta\rho<2.5\cdot 10^{-3}$, while a dashed line corresponds to $\Delta\rho<5\cdot 10^{-3}$. We notice that the three-site model is already a very good approximation for the continuum model, with a difference of just 3\%.}\label{fig:mchi}

Current bounds (see for example Langacker and Erler~\cite{Eidelman:2004wy}) yield approximately $\Delta\rho<2.5\cdot 10^{-3}$, at 90\% CL, assuming a moderately heavy (340 GeV) Higgs boson, and $\Delta\rho<5\cdot 10^{-3}$ in the case of a heavy (1000 GeV) Higgs boson. In Fig.~5 (left) we show the corresponding lower bounds on $m_{\chi_1}$ as a function of $m_{W_1}$, with $N$ varying from 1 to 122. In Fig.~5 (right) we plot the lower bounds on $m_{\chi_1}$ as a function of $N$, for $m_{W_1}$=500 GeV. In each case we add $(3\alpha/16\pi c^2)\log(m_{W_1}^2/(m_H^\textrm {ref})^2)$ to the experimental upper bound on $\Delta\rho$, in order to take into account the small hierarchy between $m_{W_1}$ and $m_H^\textrm {ref}$: this gives an appreciably weaker bound for $m_{\chi_1}$ only for $m_H^\textrm {ref}$=340 GeV. We notice that the three-site model is already a very good approximation for the continuum limit: the 17\% difference in $f(N)$ between $N=1$ and $N\rightarrow\infty$ is reduced to approximately 3\% for $m_{\chi_1}$. This is because the factor in front of $\sqrt{f(N)}$, in Eq.~(\ref{eq:bound}), behaves approximately as the inverse of $\sqrt{f(N)}$ itself, for relatively large values of $N$. We also notice that, for values of $m_{W_1}$ within the unitarity bounds, the fermion mass scale is approximately one order of magnitude larger than the gauge mass scale~\cite{SekharChivukula:2006cg}. This implies that even the lightest among the heavy fermions is probably well beyond the reach of LHC.

\section{Conclusions}\label{sec:conclusions}
The most relevant aspect of Higgsless models, either from extra-dimensions or from deconstruction, lies in the low energy behavior, since the light fermion profiles can be adjusted to minimize the impact on the elctroweak observables, without pushing the new-physics scale above the bounds imposed by unitarity. In the models we analyzed, the leading order tree-level corrections to the electroweak parameters are proportional to the quantities which determine the amount of delocalization of the standard model gauge bosons, $x^2$, and the left-handed light fermions, $\varepsilon_L^2$. The right-handed light fermions are virtually exactly localized, since the quantity parametrizing delocalization for a flavor $\chi$, $\varepsilon_{\chi_R}^2$ is related to the mass of the lightest mode by $m_{\chi_0}=M\ \varepsilon_L\ \varepsilon_{\chi_R}$, with $M$, the fermion mass scale, and $\varepsilon_L$ being universal. Then $m_{\chi_0}=0$ requires $\varepsilon_{\chi_R}=0$. If $x^2$ and $\varepsilon_L^2$ are related like in Eq.~(\ref{eq:Yahoo-delocal}), then to leading order the $S$ parameter vanishes. $T$ and $U$ are naturally zero to order ${\cal O}(x^2,\varepsilon_L^2)$, because of custodial isospin. Moreover, since $x^2$ is very small, $m_W^2$ is much smaller than $m_{W_1}^2$, and thus to leading order the $\rho$ parameter is the same as $1+\alpha T$. In summary we can therefore say that the natural sizes for the tree-level correction to the electroweak observables are  $x^2$ and $\varepsilon_L^2$, but the overall coefficients are zero either by symmetry (like for $\Delta\rho$, $T$, and $U$), or by fine tuning (like for $S$).

The natural size for the one-loop corrections is $g^2/16\pi^2$, and this can still be large. In this paper we computed the fermionic one-loop correction to the $\rho$ parameter in the SU(2)$\times$SU(2)$^N\times$U(1) Higgsless model, with a flat background for SU(2)$^N$, and for $N$ varying from 1 to $\infty$. We focused on the new-physics contribution only, which is essentially due to loops with top and bottom heavy modes (with at least one heavy mode running in the loop). In fact in our Higgsless model the violation of custodial isospin is encoded in different values of $\varepsilon_{\chi_R}$ within a single doublet. But light fermions have $\varepsilon_{\chi_R}\simeq 0$, and thus don't contribute, while in the top-bottom doublet $\varepsilon_{t_R}$ is certainly non negligible. We then argued that the leading order new-physics contribution to $\Delta\rho$ is given by setting $\varepsilon_L=0$, and has the form shown in Eq.~(\ref{eq:NDr}), for an arbitrary-$N$ model. Our results show that, as far as the contribution to $\Delta\rho$ is concerned, the three-site model ($N=1$) is already an excellent approximation for the continuum model ($N\rightarrow\infty$). This is best seen from the lower bounds on the fermion mass scale, which arise from the experimental upper bounds on $\Delta\rho$: The dependence on $N$ is very weak, and the difference between the three-site model and the continuum model turns out to be just 3\%. In agreement with the results of Ref.~\cite{SekharChivukula:2006cg} we find that the mass of the lightest among the heavy fermions is at least $\sim 10^1$ times larger than the mass of the $W_1$ boson, and therefore beyond the reach of LHC.

\centerline{\bf Acknowledgments}
We thank Sekhar Chivukula and Elizabeth Simmons for illuminating discussions and advice. We also thank Carl Schmidt for useful discussions and comments. B.C. and S.D.C. are supported in part by the US National Science Foundation under grant PHY-0354226.  R.F. is supported by the Marie Curie Excellence Grant under contract MEXT-CT-2004-013510. 


\end{document}